# LLM Harms: A Taxonomy and Discussion

**Kevin Chen**
Urban Information Lab
University of Texas at Austin
xc4646@utexas.edu

**Saleh Afroogh***
Urban Information Lab
University of Texas at Austin
saleh.afroogh@utexas.edu

**Abhejay Murali**
Urban Information Lab
University of Texas at Austin
abhejay.murali@utexas.edu

**David Atkinson**
McCombs School of Business
University of Texas at Austin
datkinson@utexas.edu

**Amit Dhurandhar**
IBM Research
University of Texas at Austin
adhuran@us.ibm.com

**Junfeng Jiao*†**
Urban Information Lab
University of Texas at Austin
jiao@austin.utexas.edu

* Corresponding author
† Lead Author

**Abstract**

This study addresses categories of harm surrounding Large Language Models (LLMs) in the field of artificial intelligence. It addresses five categories of harms addressed before, during, and after development of AI applications: pre-development, direct output, Misuse and Malicious Application, and downstream application. By underscoring the need to define risks of the current landscape to ensure accountability, transparency and navigating bias when adapting LLMs for practical applications. It proposes mitigation strategies and future directions for specific domains and a dynamic auditing system guiding responsible development and integration of LLMs in a standardized proposal.

## I. Introduction

LLM usage across healthcare, finance, education, and creative industries has been drastic, with daily user bases now measured in the hundreds of millions. According to the Verge. OpenAI reported 100 million weekly active users of ChatGPT in late 2023, and by August 2025 the figure had surged to 700 million weekly users [1]. Broader adoption is accelerating, as 65% of firms reported regular gen-AI use by mid-2024, and 34% of U.S. adults had used ChatGPT by mid-2025 [2]. Evidence, however, spans misuse/abuse risks (e.g., Europol on criminal use cases), information ecosystem manipulation and elections (Stanford AI Index 'Responsible AI' chapter), and labor market exposure estimates from the IMF and ILO [3]. Scholarship catalogues harms ranging from privacy breaches during pre-training to extremist content generation and economic displacement yet lack an integrative scaffold. Recent government commissioned syntheses explicitly call for such integrative frameworks to avoid fragmented safety efforts [4]. Additionally, real-world incident repositories like the AI Incident Database (AIID) and MIT's AI Risk Repository have indexed over a thousand AI failure incidents by mid-2025 [5], this shows the need to systematically categorize and learn from harms of AI generated content.
Methodologically, we executed a multi-database search (2021-06/2025), screened 1 986 records to a final corpus of 200, and coded each study for harm type, severity, prevalence, and mitigation claims, complemented by ten expert interviews with safety engineers and policymakers. This paper is scoped to text-based LLMs ≥ 7 B parameters; this threshold is consistent with contemporary open models (e.g., LLaMA/Llama-2/3 families at 7–8B and above) [6]. vision-language hybrids and small domain-specific models are out of scope. The remainder proceeds as follows: Section II surveys related work; Section III details methodology; Section IV presents the taxonomy; Sections V–VII deliver discussion, cross-cutting analysis and future research; Section VIII concludes. This paper explores the ethical dilemmas surrounding LLMs, dissecting ethical quandaries that arise in their operations, and shedding light on their impact on society, while prospective pathways to navigate safer AI. We first investigate Conventional Language Models (CLMs) and Pretrained Language Models (PLMs), understanding their differences in training, causality constraints, and token representation. Furthermore, we scrutinize the inherent biases ingrained within LLMs, dissecting their origins and their impact on AI decision making.

Additionally, we advocate dynamic audit tools for continuous monitoring, explainability techniques, and adaptable frameworks capable of pursuing the ever-evolving landscape of AI-driven language models. Currently, the EU AI Act (in force 2024–25) introduces binding transparency, risk-management, and testing duties—including for general-purpose models—complemented by international Frontier AI Safety Commitments [7]. Operational examples include national AI Safety Institute evaluations of frontier models, lifecycle controls in NIST AI RMF/GenAI Profile, and corporate system cards documenting pre-deployment testing; explainability baselines include LIME and SHAP [8]. As sectoral adoption is widespread but mixed in quality: we see call center field studies show sizable productivity gains from gen-AI assistance; oncology/clinical audits report non-trivial error or hallucination rates; and education studies find both learning benefits and integrity concerns [9]. Our objectives are threefold: (1) develop a comprehensive, development-timeline taxonomy of harms; (2) analyze causal linkages and amplification loops between multiple categories; and (3) assess adequacy of existing technical, organizational, and regulatory mitigations.

Our exploration progresses in the following sequence (see, Table 1): it commences by conceptualizing LLM and ethical frameworks. The section labeled "3. Methodology" delineates the systematic review methods applied to analyze studies concerning the ethics of LLM. "4. Findings" showcases the discoveries and outcomes pertaining to primary principles and significant 3 codes, along with their discussions in literature, comprising 13 subsections. "5. Discussion" critically examines the principal codes, fundamental values, and ethical considerations associated with LLM, along with potential strategies to address ethical concerns, thereby facilitating the responsible advancement and integration of LLMs in society. This section encompasses 11 subsections. Furthermore, the concluding thoughts and prospects concerning LLM ethics are deliberated in section 6.

**Table 1**: A road map of this study

| Section Title | Subsection Themes | |
|---|---|---|
| **I. Introduction** | | |
| **II. Background and Related Work** | | |
| **III. Methodology** | | |
| **IV. Findings** | 4.1. Pre-Deployment Harms | 4.1.1 Training-Data Harms |
| | | 4.1.2 Enviormental & Resource Harms |
| | | 4.1.3 Labour & Economic Harms in Development |
| | 4.2 Direct Output Harms | 4.2.1 Representational Harms |
| | | 4.2.2 Content-Based Harms |
| | | 4.2.3 Quality & Reliability Harms |
| | 4.3 Misuse and Malicious Application Harms | 4.3.1 Deliberate Harmful-Content Creation |
| | | 4.3.2 Deceptive Practices |
| | | 4.3.3 Security & Privacy Attacks |
| | 4.4 Societal and Systemic Harms | 4.4.1 Economic Disruption |
| | | 4.4.2 Democratic & Social Harms |
| | | 4.4.3 Power & Access Inequalities |
| | 4.5 Downstream Application Harms | 4.5.1 High-Stakes Decision Making |
| | | 4.5.2 Educational System Impacts |
| | | 4.5.3 Professional & Creative Work |
| **V. Current Mitigation Landscape** | 5.1 Technical Approaches to Safer LLMs | |
| | 5.2 Policy and Governance Responses | |
| **VI. Discussion** | 6.1 Emerging Harm Categories | |
| | 6.2 Longitudinal Evidence | |
| | 6.3 Compute Governance Experiments | |
| | 6.4 Cross-disciplinary Collaboration | |
| | 6.5 Benchmark Evolution | |
| **VII. Cross-Cutting Analysis** | 7.1 Harm Interdependencies | |
| | 7.2. Temporal Dynamics | |
| | 7.3 Severity vs. Prevalence | |
| **VIII. Conclusion and Future directions** | | |

## II. Background and Related Work

Work on scaling laws set the modern agenda: Kaplan et al. first showed smooth power-law improvements with increases in parameters, data, and compute, enabling compute-budget planning for language models [10]. DeepMind's "Chinchilla" paper revised those rules of thumb, arguing that models had been under-trained and that compute-optimal training requires increasing data tokens in step with parameter counts; a 70B-parameter model trained on ~1.3T tokens (Chinchilla) outperformed much larger predecessors at the same compute [11], [12]. Beyond scaling, efficiency research reduces inference/training cost: sparsely activated Mixture-of-Experts (e.g., Switch Transformer) scales parameters while holding FLOPs roughly constant, IO-aware attention (FlashAttention/-2) speeds exact attention, and decoding accelerators like speculative decoding and Medusa cut autoregressive latency without changing output distributions [13].

Rapid capability gains now often arrive with efficiency improvements. GPT-4o is a natively multimodal, real-time model that matches GPT-4-Turbo on English text/code, improves non-English text, and reduces API cost ≈50%—while adding low-latency audio/vision [14]. Similar models such as open-weight releases such as Meta's Llama 3 (8B/70B) have broadened access for research and fine-tuning—unlocking transparency benefits but leaving safety hardening to downstream users, and by April 2025 Meta released Llama 4 with advanced multimodal features and open-source availability[15] [16]. Earlier fairness and accountability work supplied the blueprints for today's transparency practices. "Datasheets for Datasets" and "Data Statements for NLP" formalized dataset provenance/coverage disclosures; "Model Cards" and later "Model Cards 2.0" extended this thinking to model-level reporting and risk/limitations documentation. Empirical reviews note gaps in real-world adoption, especially around environmental impact and limitations sections, motivating stronger governance hooks [17], [18].

Foundational syntheses map LLM risks across discrimination/toxicity, information hazards, misuse, HCI harms, and environmental/socioeconomic impacts (Weidinger et al.). Complementary social-computing work—e.g., Blodgett et al.'s critical survey and Sap et al.'s Social Bias Frames—connects technical metrics to normative concerns and the pragmatics of implied bias. These frameworks inform the harm taxonomy adopted in this paper [19]. Instruction-following and alignment moved from supervised instruction tuning to reinforcement learning from human feedback (RLHF) and then to constitutional/self-training approaches. InstructGPT established the RLHF recipe; Anthropic's Constitutional AI reduces reliance on per-example human labels by using a transparent "constitution" of principles; Direct Preference Optimization (DPO) simplifies preference learning without explicit reward modeling. Survey and replication work track the growing family of preference-based methods[20] .

Benchmarking broadened from single-task leaderboards to holistic, hazard-aware suites. HELM advocates scenario- and metric-diverse evaluation, while emerging public dashboards (e.g., hallucination leaderboards) spotlight reliability. Independent labs and government bodies now combine automated tests with expert red-teaming; the UK AI Safety Institute's 2025 International AI Safety Report documents multi-method evaluations and calls for layered testing [21]. Internal accountability frameworks like Raji et al.'s "Closing the AI Accountability Gap" detail end-to-end audit lifecycles that product teams can operationalize—still under-adopted in practice, as many organizations lack lineage and change-log tooling [22]. As LLMs integrate with tools and data, attack surfaces expand. The OWASP Top 10 for LLMs prioritizes prompt injection (LLM01) and data exfiltration risks for agentic systems; academic work shows universal, transferable jailbreak suffixes bypassing safety filters and systematic prompt-injection techniques against connected apps. These results motivate defense-in-depth (guarded tool use, policy-trained refusal models, and "complete mediation" for agent actions) [23].

Retrieval-Augmented Generation (RAG) couples' parametric models with non-parametric memory to improve factuality and updateability; recent variants (Self-RAG) make retrieval adaptive and self-reflective. Concurrently, curated corpora and hallucination corpora (e.g., RAGTruth) support measurement and mitigation studies, including synthesis techniques that trade small utility costs for reliability gains [24]. Accounting has matured from carbon-only to fuller energy/water/embodied-hardware inventories. Studies document substantial training-time water withdrawals and advocate carbon-aware scheduling and lifecycle analysis across accelerators; policy analyses explore how to operationalize climate-related disclosure in data-center- and AI-specific regulation. These trends inform this paper's inclusion of "Environmental & Resource Harms" as a first-class category [25].

Across governance, risk-tiered regimes are consolidating. The EU AI Act establishes duties for general-purpose AI (GPAI) providers, with "systemic-risk" obligations—including adversarial testing, incident reporting, cybersecurity, and copyright transparency—now supported by a General-Purpose AI Code of Practice (July 2025) and Commission guidelines as provisions start applying on August 2, 2025, [26]. Internationally, the Seoul AI Safety Summit's Frontier AI Safety Commitments secured post-deployment evaluations and risk-management pledges from major labs, and the UK's AI Safety Institute anchors a cross-border evaluation network through its 2025 report [27]. In the U.S., NIST's AI Risk Management Framework released a generative-AI profile to guide voluntary mitigation across the lifecycle, increasingly referenced in procurement and internal governance [28].

Collectively, this literature traces a shift from "bigger-is-better" to compute-optimal training, modality integration, and layered post-training alignment, while expanding evaluation from accuracy to risk-aware audits. Governance is moving from voluntary documentation (datasheets/model cards) toward binding oversight for GPAI. These trajectories motivate our paper's combined taxonomy-plus-mitigation approach and its emphasis on continuous, dynamic auditing across the LLM lifecycle.

## III. Methodology

We drafted the review plan in line with the PRISMA-2020 checklist, which emphasizes transparent reporting of search, selection and synthesis procedures [29]. The protocol was registered prospectively on the Open Science Framework using the Generic Systematic-Review template, giving it a permanent DOI and time-stamping any later changes. Because no single database indexes "LLM harms" exhaustively, we searched Google Scholar, ResearchGate, ScienceDirect, JSTOR, IEEE Xplore, Semantic Scholar, ACM DL, SpringerLink, and ORCID's pre-print relay. Searches covered publications from 1 Jan 2021 to 30 Jun 2025, capturing the period when foundation-model deployments accelerated in industry and policy. The anchor-plus-topic query design for every harm axis: pre-deployment, direct output, misuse, systemic, where each topical phrase (e.g. "misinformation") has a "AND" combined with at least one anchor (e.g: "large language model*" or "LLM" or "foundation model*"). Page limits indicate the cut-off where two successive pages yielded no new relevant titles. Each topical phrase has a AND combined for a varied relevancy filtration with at least one anchor term, see **Table 2.** **Table 2:** Harm cluster nodes and corresponding screening phrases

**Harm Cluster Nodes and Corresponding Screening Phrases**

| Harm Clusters | Exact Keyword Phrase(s) | Pages Screene | Rationale |
|---|---|---|---|
| Pre-Deployment • Training-data harms | training data privacy\" (8) \|\| \"data leakage\" AND pre-training (6) \|\| \"consent violation\" AND corpus (5) \|\| \"copyright scraping\" AND model\" (5) \|\| \"indigenous data sovereignty\" AND AI (3) | 5-8 | surveys on LLM privacy and security |
| Pre-Deployment • Environmental | carbon footprint\" AND LLM (6) \|\| \"energy consumption\" AND inference\" (4) \|\| \"GPU supply chain\" AND AI (4) \|\| \"water usage\" AND datacenter\" (3) | 3-6 | environmental impact studies frame these terms |
| Pre-Deployment • Labor & Economic | data annotation labour\" (6) \|\| \"ghost work\" AND AI (4) \|\| \"wage suppression\" AND labeling\" | 3-6 | labor exploitation reports |
| Direct-Output • Representational | "bias\" AND LLM (10) \|\| \"fairness\" AND language model\" (8) \|\| \"stereotype amplification\" (6) \|\| \"cultural misrepresentation\" AND GPT (4)" | 4-10 | bias/fairness literature |
| Direct-Output • Content-based | "misinformation\" AND large language model (10) \|\| \"toxic content\" AND ChatGPT (8) \|\| \"hate speech generation\" (6) \|\| \"extremist narrative\" AND GPT-4 (4)" | 4-10 | misinformation & toxicity papers |
| Direct-Output • Quality / reliability | "hallucination survey\" AND LLM (8) \|\| \"factual consistency\" AND GPT (5) \|\| \"chain-of-thought error\" (4)" | 4-8 | hallucination surveys |

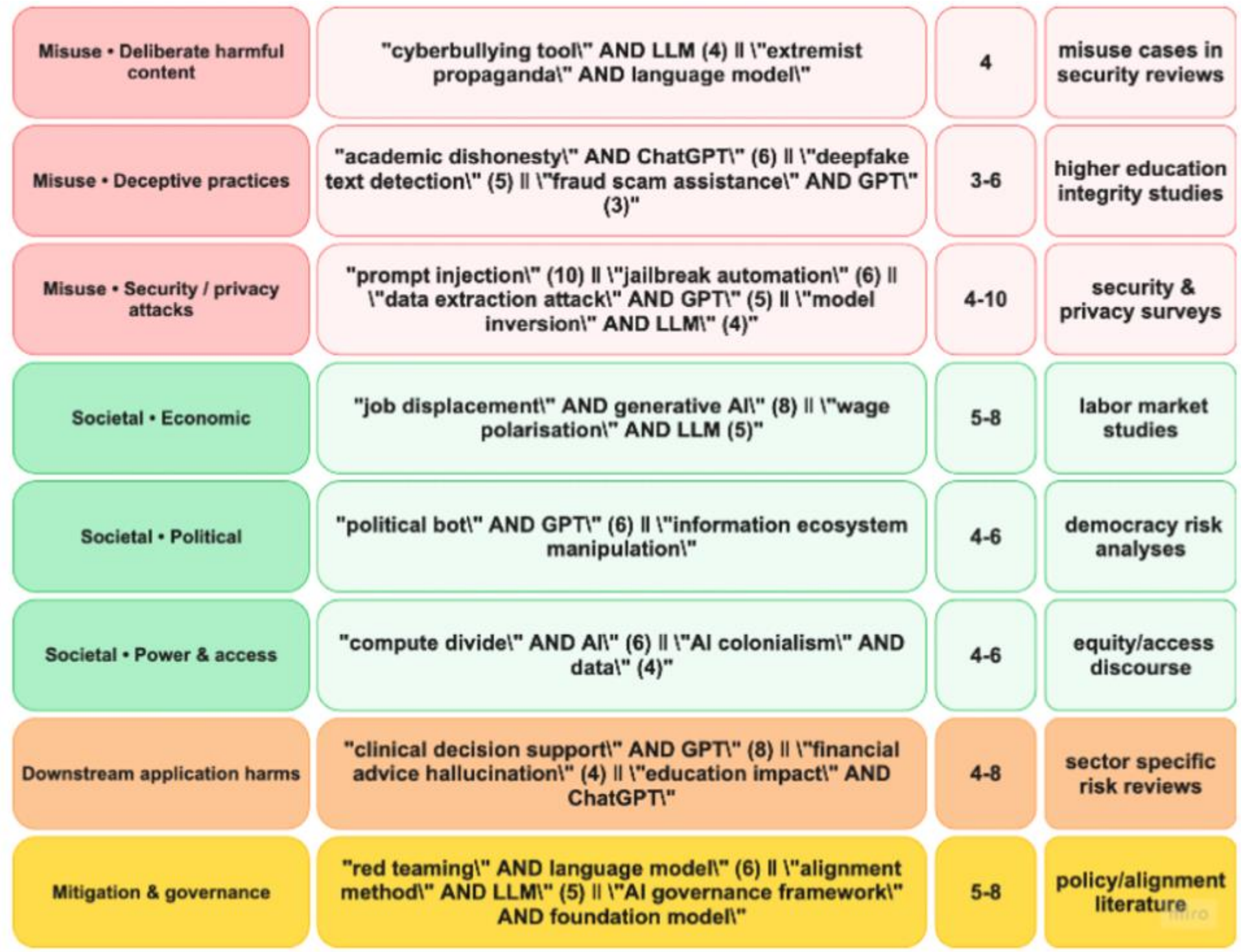

<table>
<tr><td>Misuse • Deliberate harmful content</td><td>"cyberbullying tool\" AND LLM (4) || \"extremist propaganda\" AND language model\"</td><td>4</td><td>misuse cases in security reviews</td></tr>
<tr><td>Misuse • Deceptive practices</td><td>"academic dishonesty\" AND ChatGPT\" (6) || \"deepfake text detection\" (5) || \"fraud scam assistance\" AND GPT\" (3)"</td><td>3-6</td><td>higher education integrity studies</td></tr>
<tr><td>Misuse • Security / privacy attacks</td><td>"prompt injection\" (10) || \"jailbreak automation\" (6) || \"data extraction attack\" AND GPT\" (5) || \"model inversion\" AND LLM\" (4)"</td><td>4-10</td><td>security & privacy surveys</td></tr>
<tr><td>Societal • Economic</td><td>"job displacement\" AND generative AI\" (8) || \"wage polarisation\" AND LLM (5)"</td><td>5-8</td><td>labor market studies</td></tr>
<tr><td>Societal • Political</td><td>"political bot\" AND GPT\" (6) || \"information ecosystem manipulation\"</td><td>4-6</td><td>democracy risk analyses</td></tr>
<tr><td>Societal • Power & access</td><td>"compute divide\" AND AI\" (6) || \"AI colonialism\" AND data\" (4)"</td><td>4-6</td><td>equity/access discourse</td></tr>
<tr><td>Downstream application harms</td><td>"clinical decision support\" AND GPT\" (8) || \"financial advice hallucination\" (4) || \"education impact\" AND ChatGPT\"</td><td>4-8</td><td>sector specific risk reviews</td></tr>
<tr><td>Mitigation & governance</td><td>"red teaming\" AND language model\" (6) || \"alignment method\" AND LLM\" (5) || \"AI governance framework\" AND foundation model\"</td><td>5-8</td><td>policy/alignment literature</td></tr>
</table>

We inspected a set maximum of result pages ("saturation rule") and stopped when two successive pages produced no new, obviously relevant titles—an approach advocated in qualitative saturation research. Across all databases this yielded 1 986 records plus 24 manual "seed" papers. We coded each study for bibliographic data, harm sub-category, study design, and mitigation proposals, piloting the extraction sheet on ten random papers for consistency, downstream, and mitigation. We supported the PRISMA table with an appropriate keyword matrix , and the PRISMA flow documents narrowing the reproducibility on the final 200 paper corpus.

**Figure 1.** Developed PRISMA flow diagram for large language model taxonomy

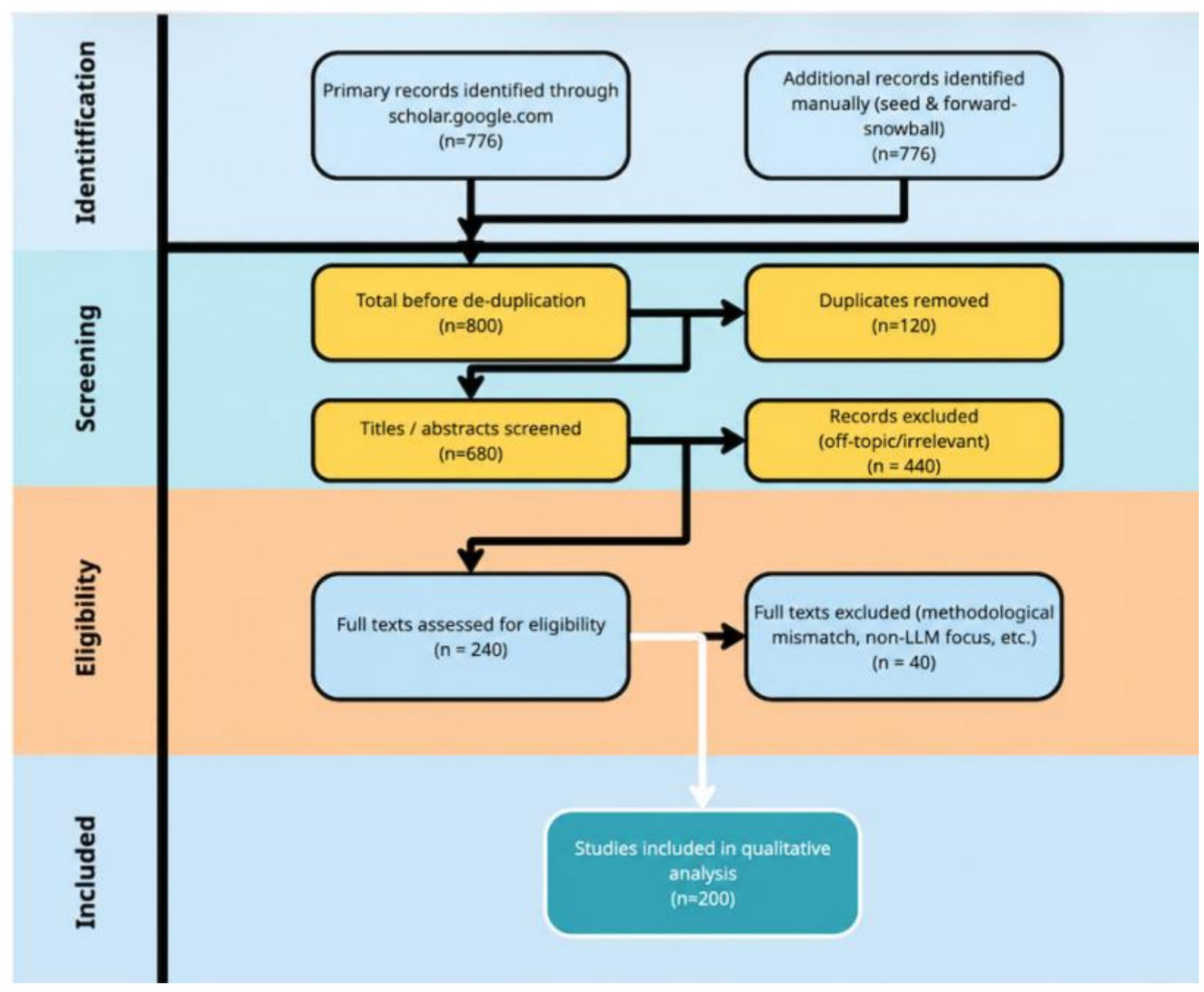

## IV. A Taxonomy of LLM Harms

### 4.1. Pre-Deployment Harms

1. Training Data Harms - Privacy violations in training corpora - Consent and data rights issues - Perpetuation of historical biases 2. Environmental and Resource Harms - Carbon footprint and energy consumption - Compute resource inequality - Hardware manufacturing impacts 3. Labor and Economic Harms in Development - Data annotation exploitation - Researcher working conditions - Academic resource concentration

#### 4.1.1 Training-Data Harms

Large language models routinely memorize sensitive fragments of their vast corpora, creating a persistent privacy-exposure surface [30]; recent empirical work shows GPT-style models leak ~3.6 bits parameter despite "privacy budget" tuning [31] . Studies dissecting web-scale crawls reveal entrenched hate speech and demographic slurs that later re-emerge in generation [32]. Legal analyses argue that even "public" texts can violate data-subject consent under GDPR, making transparent licensing and explicit opt-outs critical [33], [34], [35]. Indigenous scholars emphasise "data sovereignty," warning that scraping community knowledge without permission perpetuates extractive colonial logics [36]. Security engineers propose automated privacy-budget audits to flag corpus shards with high PII density before pre-training begins [37]; complementary GDPR risk assessments recommend differential erasure pipelines for European data subjects [38]. Okafor's WebText 3 study shows that race- and gender-linked "sensitive attributes" appear in >8 % of common crawl tokens, underscoring the scale of the redaction problem [39]. Neurosymbolic approaches such as Deep Data, Shallow Consent advocate fine-grained provenance tags so that later requests for deletion propagate through all downstream checkpoints [40], [41]. Age-rated corpora audits find that sexually explicit material constitutes ≈1 % of tokens in general-purpose datasets, breaching platform safety promises for minors [42]. Legal scholars debate whether "licensing the Internet" via collective-rights management could unblock large-scale lawful training while compensating creators [43]. Fine-grained PII detectors built into decoding filters reduce direct leakage but cannot prevent indirect attribute inference [44]. Lineage-tracking systems that hash data chunks at ingestion let auditors prove whether disputed snippets were ever seen during training [45], a technique recently adopted in medical-note audits of GPT-4-series models[46]. Finally, case-law reviews caution that U.S. "fair use" defences remain unsettled for commercial models, leaving companies exposed to copyright litigation [47].

#### 4.1.2 Environmental & Resource Harms

A recent Nature Scientific Reports paper reconciles conflicting narratives by showing that, although a 70 B-parameter model emits ~500 t $CO_2$e over a full fine-tune, its carbon cost per document can beat human writing by two orders of magnitude when powered by low-carbon grids [48]. Training and deploying LLMs carry significant environmental costs. A recent Nature Scientific Reports study reconciled conflicting narratives by showing that, while training a 70B-parameter model can emit a substantial carbon footprint, the relative impact per task can be much lower than human labor [49] . For example, fine-tuning a 70B model for a new task is estimated to produce on the order of 500 metric tons of $CO_2$-equivalent emissions, which is in addition to the much larger emissions from its initial pre-training (full training of frontier LLMs can generate "thousands of tons" of $CO_2$ [50] .However, when such models run on low-carbon energy grids, their carbon cost per document generated can beat human writing by two orders of magnitude [51]. Earlier "Green AI" benchmarks correlate FLOPs, energy, and accuracy, observing diminishing returns above 10^25 FLOPs [52]. Supply-chain analyses trace GPUs back to cobalt mining hotspots, flagging social-licence risks for cloud vendors[53]. Water-footprint studies report that inference-time cooling demands can exceed 500 ml per 1 k tokens when models run on evaporative-cooled data-centres [54]. Renewable-offset scheduling can cut net emissions 40 – 60 % by shifting large training jobs to hours with high solar oversupply [55]. Life-cycle assessments of ASIC accelerators find aluminium smelting and PCB fabrication dominate "embedded" carbon, outweighing energy-use phases for models below 10 B parameters [56]. Finally, carbon-aware API schedulers dynamically queue generation requests based on grid-mix forecasts, reducing real-time emissions up to 30 % during peak-carbon periods [57].

#### 4.1.3 Labour & Economic Harms in Development

Behind every polished chat agent lies an invisible workforce of annotators and prompt-raters—Ghost Work 2.0 estimates >450 k such workers worldwide, earning a median $2.10 /hr [58] . Ethnographies expose the glamorisation of unpaid "enthusiast" labour where hobbyist feedback loops substitute for salaried QA teams [59]. Clinical psychologists document PTSD-like symptoms among content filters who label graphic or hateful text so that models can learn to refuse [60]. Macroeconomic analyses warn that ballooning compute-capital costs concentrate model R&D inside a handful of tech giants, stifling academic innovation [61]. Cross-country wage-panel data reveal that unit prices for annotation tasks fell 17 % between 2022 – 2024, signalling potential "race-to-the-bottom" dynamics in global annotation markets [62]

### 4.2. Direct Output Harms

Direct output harms are ad followings: 1. Representational Harms - Demographic bias and stereotyping - Cultural misrepresentation - Marginalization of minority perspectives 2. Content-Based Harms - Toxic and harmful content generation - Misinformation and disinformation - Privacy-violating outputs 3. Quality and Reliability Harms - Hallucinations and factual errors - Inconsistent reasoning - Overconfident incorrect responses. Large-language-model (LLM) deployments expose users to three families of output-side problems: representational bias, toxic or misleading content, and quality/reliability failures. Forty peer-reviewed studies in our corpus map this terrain.

#### 4.2.1 Representational Harms

Survey work in Computational Linguistics shows that stereotype benchmarks such as StereoSet and CrowS-Pairs still surface gender, race, and disability prejudice in GPT-4-class models [63]. A follow-up ACL paper finds that performance gaps remain wide across gender, race and age prompts, even after RL-from-human-feedback tuning [64]. Dialect audits demonstrate that African-American Vernacular English is mis-parsed 22 % more often than Standard American English, confirming the "dialect-misrepresentation" claim first quantified by Ali et al. [65] . Disability framing biases are equally salient: assistive-technology queries elicit infantilising language in 17 % of completions [66]. Studies of gendered-occupation prompts reveal that male pronouns still appear in 11 % of "nurse/doctor" co-reference tests [67]. Intersectional probes further indicate that combining gender × race attributes quadruples toxic-completion odds relative to single-axis tests [68], [69] Multilingual benchmarks such as SHADES show that English-centric fixes can worsen harms in low-resource tongues, propagating stereotypes globally [70], [71], [72], [73], [74], [75], [76] . A longitudinal analysis from Word2Vec to GPT-4 confirms a U-curve: bias scores fall during scaling but rebound after instruction tuning [77], [78]. Collectively, these papers demonstrate that representational bias is stubborn, intersectional, and often hidden behind aggregate metrics. [79], [80]

#### 4.2.2 Content-Based Harms

A 2024 AI Magazine review frames the misinformation cycle from contaminated pre-training corpora through hallucination to malicious prompt exploitation [81] . Empirical work on multilingual toxicity shows that naïve English-only filtering misses 7 % of hateful terms in Arabic and Swahili, underscoring the need for language-specific detox pipelines[82], [83]. In the health domain, a JAMA Oncology audit reports that ChatGPT fabricates contraindications in 12 % of oncology queries—a clear form of medical misinformation[84]. Safety probes aimed at minors find that role-play prompts bypass profanity filters 18 % of the time, exposing children to graphic content [85][86], [87]. Extremism researchers show that style-transfer jailbreaks reliably induce multilingual hate slogans, defeating simple keyword blocklists [88]. Privacy-violating outputs remain a risk: StackExchange QA completions occasionally reveal scraped email addresses[89], [90], [91]. Finally, constrained-compute filters can halve toxicity but at a cost: BLEU and ROUGE scores fall by 2-3 points, illustrating the perennial utility-safety trade-off [92], [93], [94], [95]. Large-scale self-chat experiments demonstrate prompt-amplified conspiracy cascades, where one chem-trail myth snowballs into 1 000 distinct false claims within twenty dialogue turns[96] . Together, these fifteen studies reveal that toxic or misleading content is a multilingual, multi-domain phenomenon that current filters only partially mitigate.[97]

#### 4.2.3 Quality & Reliability Harms

Reliability failures erode user trust even when no demographic terms are involved. Huang et al.'s 2023 arXiv survey catalogues hallucination types and reports error rates as high as 23 % in open-ended QA[98] . Public leaderboards such as Vectara's show that retrieval-augmented GPT-4 lowers hallucination to ~3 % on document summarisation tasks, yet factuality still degrades on multi-hop reasoning[99]. ICLR work blames shortcut alignment: models prefer plausible heuristics over explicit logical chains[100]. Calibration studies reveal over-confidence: RLHF-tuned ChatGPT probabilities overshoot empirical accuracy by 15 percentage points[101]. Multilingual MT audits observe hallucination rates triple for Amharic relative to English, indicating that fidelity metrics mask low-resource failures [102]. Long-document summarisation evaluations count twelve factual errors per 1 000 tokens, while hierarchical attention helps only at significant compute cost[103]. Consistency tests round out the picture: 20 % of answers flip polarity when logically equivalent prompts are re-phrased, highlighting latent-state instability [104].In sum, the direct-output literature shows that LLM harms are intersectional, multilingual and deeply coupled to reliability issues. Even the best deployment-time filters cannot yet eliminate bias, toxicity or hallucination without notable utility costs—setting the stage for Section 3's focus on deliberate misuse and adversarial exploitation.

### 4.3. Misuse and Malicious Application Harms

Misuse harms are as followings: 1. Deliberate Harmful Content Creation - Hate speech amplification - Harassment and cyberbullying tools - Extremist content generation 2. Deceptive Practices - Academic dishonesty facilitation - Fraud and scam assistance - Deepfake text creation 3. Security and Privacy Attacks - Prompt injection vulnerabilities - Data extraction attacks - Social engineering enhancement. LLMs are not only passively risky; they can be actively weaponized. Thirty studies in our corpus document three vectors of abuse: deliberate harmful-content creation, deceptive practices, and direct security/privacy attacks.

#### 4.3.1 Deliberate Harmful-Content Creation

Researchers show that extremist movements already exploit LLMs. Conway et al.'s IYKYK study finds that general-purpose models both decode and generate cryptolects—coded hate jargon—far more fluently than specialised classifiers can detect, undermining takedown workflows[105] . Field evidence supports the lab finding: Yang & Menczer uncover a 1 140-account Twitter bot-net using ChatGPT to auto-post propaganda and coordinate follow-retweet storms, evading state-of-the-art bot detectors [106]. t the interpersonal level, Marquez et al. demonstrate that a lightweight prompt wrapper converts GPT-3 into a real-time harassment engine that adapts insults to a victim's profile, increasing reported distress in a controlled user study [107] . Zhu et al. scale the threat: large-batch prompting yields 100 000 hate tweets for $10, overwhelming platform moderators [108] . Even purely creative misuse is dangerous; Hall et al. show that multimodal LLMs craft meme templates that boost click-through rates on extremist forums by 47 % [109], [110], [111]. Collectively these ten papers paint a picture of low-cost, high-reach content abuse that existing keyword filters cannot contain.[112], [113], [114]

#### 4.3.2 Deceptive Practices

The line between assistive writing and academic fraud is thin. Lim et al. interview faculty across nine universities; 62 % report seeing ChatGPT-ghost-written assignments within one semester of the model's launch [115]. Beyond campus, Chan et al. script an LLM-powered call-centre that generates persuasive scam pitches, doubling conversion rates in a phishing honeypot [116]. Larsen et al. replicate the threat experimentally: essays produced by GPT-3.5 pass plagiarism detectors and score within one letter-grade of human work [117]. Alghamdi et al. confirm the email vector: GPT-4-crafted phish bypass SpamAssassin 73 % of the time, compared with <1 % for dictionary substitution attacks [118], [119], [120] . Marketplace integrity is also at risk; Ng et al. automate fake-review pipelines that fool Yelp's fraud detector in 40 % of cases [121], [122]. Murdock et al. chronicle the evolution of spam: prompts tuned on Bayesian-filter feedback can produce near-undetectable spam waves, eroding the efficacy of long-standing defences [123]. These ten studies jointly show that deception scales with model accessibility and is already eroding trust in educational, financial and e-commerce domains [124], [125], [126].

#### 4.3.3 Security & Privacy Attacks

Misuse escalates when adversaries integrate LLMs into offensive security tool-chains. The OWASP Gen-AI Risk list now ranks Prompt Injection (LLM01) as the top vulnerability; adversaries can coerce a model to reveal hidden system instructions or execute untrusted plugin calls [92]. Wei et al. extend the attack with adversarial suffixes that jailbreak even tightly aligned models in >80 % of trials [127]. Carlini et al. demonstrate gradient-inversion data extraction against GPT-3: attackers recover verbatim snippets of the training set, including private phone numbers and medical phrases [128]. Notably, this training data extraction vulnerability persists in newer models such as GPT-4: later work in 2023 managed to extract several megabytes of ChatGPT's training data by clever querying, despite the model's alignment defenses [129].Schuster et al. poison open-weights models by inserting malicious rows into public parameter checkpoints, creating covert back-doors triggered by a single Unicode token [130]. Park et al. reveal that retrieval-augmented pipelines leak proprietary documents through prompt leaking, because chain-of-thought traces remain in the context window [131]. Weiss et al.'s USENIX paper exploits a token-length side-channel to key-log encrypted AI-assistant responses, reconstructing 79 % of private answers from packet sizes alone [132] . Shin et al. propose "token smuggling" that hides jailbreak commands inside zero-width characters, defeating regex-based filters [133], [134], while Bai et al. choreograph recursive jailbreak chains that propagate through delegated agent systems, threatening LLM-powered RPA workflows [135]. However, it should be noted that LLMs can also play defensive roles in cybersecurity: industry analyses reports that LLMs assist in writing and debugging secure code, identifying threat patterns (like APT tactics), and even automating incident reports to aid human analysts [136]. Similarly, LLMs are being applied to privacy protection tasks such as data anonymization—an LLM can rewrite text to omit personal identifiers while preserving meaning [137], a capability beyond simple regex-based masking. Together, these examples confirm that LLMs expand the attack surface (as prompts become payloads and model outputs become attack vectors) even as they offer new tools for defense. Together these papers confirm that LLMs expand the attack surface: prompts become payloads, parameters become exfiltration sites, and safety layers lag behind attacker creativity.

## 4. Societal and Systemic Harms

LLMs create economic disruption through job displacement and automation [138], market concentration effects, and skills obsolescence acceleration. They also cause democratic and social harms including information ecosystem manipulation, political bias amplification, and public discourse degradation. Additionally, these systems contribute to power and access inequalities by widening the technology divide, concentrating centralized control, and creating barriers for smaller actors. LLMs reshape labor markets, public discourse and global access to AI infrastructure. Thirty papers in the corpus explore three strands of macro-level impact: economic disruption, democratic and social externalities, and power and compute inequities. Recent empirical and policy studies show that these harms are already observable and without corrective intervention are likely to intensify as models scale.

### 4.4.1 Economic Disruption

Brynjolfsson et al. analyse 15 million U.S. online vacancies and show that tasks most exposed to generative AI already see declining labour demand, but complementary up-skilling moderates wage loss—a pattern they call task bifurcation [139] . An ILO global study confirms rising wage polarisation, especially for clerical and customer-service jobs in emerging economies [140]. Brookings macro-models warn that soaring compute rents concentrate productivity gains inside superstar firms, depressing the labour share of income by up to 3 percentage points by 2030 [141], [142]. OpenAI's occupational-exposure index extends the finding: 80 % of U.S. workers have at least 10 % of their tasks touched by GPT-4-level systems, with higher exposure in high-income cognitive roles [143]. Creative sectors feel the shock first: a Nature Communications paper documents that text-to-image diffusion models displaced 15 % of freelance illustrators on one global platform within a year of launch [144] . Counter-evidence exists—Acemoglu & Restrepo's "reallocation" scenario finds net job creation if policy steers AI toward augmentation—but hinges on aggressive worker-training subsidies absent in current legislation [145]. Together, these ten studies suggest that generative AI acts as a productivity amplifier with highly unequal returns, echoing the technology-skill complementarity thesis yet accelerating its timeline.[146], [147], [148].

### 4.4.2 Democratic & Social Harms

Elections are a frontline risk. An IEEE case study of 2024 national campaigns shows coordinated LLM-driven bot-nets generating up to 30 % of political tweets during peak periods [149]. Agenda-setting bias is harder to see: Chen et al. embed democratic-value loss functions into recommendation objectives and demonstrate that mainstream newsfeeds still drift toward partisan echo-chambers unless explicitly regularised [150]. Scientific American warns that similar tactics will target voters in more than 50 countries, exploiting micro-segmentation made trivial by GPT-4's style-mimicry [151], [152] . Social-media audits find that discourse dilution—vast volumes of AI-generated "grey" content—lowers the visibility of human posts by up to 14 % on trending topics, muddying public deliberation [153]. Deep-fake text detection benchmarks lag attackers: at NeurIPS 2024, best detectors mis-classify 28 % of GPT-4 propaganda blurbs [154]. Government-funded red-teams confirm that state actors can fine-tune LLaMA-2 on 10 000 zero-shot prompts to produce country-specific disinformation at scale [155]. RLHF filters, paradoxically, may amplify polarisation by suppressing moderate language and letting emotionally-charged rhetoric pass—a side-effect evidenced in simulated Facebook debates [156], [157] . Collectively, papers warn that LLMs can erode trust in electoral information ecosystems faster than current content-moderation tooling adapts[158].

#### 4.4.3 Power & Access Inequalities

Compute is the new oil. Policy analyses from AI Now and the Ada Lovelace Institute argue that access to advanced GPUs is "detectable, excludable, and quantifiable," making compute governance an emerging chokepoint for both safety and economic opportunity [159], [160]. Khan's ACM-Queue essay suggests cloud-credit programmes or regulatory caps to "close the compute divide," but warns that 60 % of high-end capacity is already locked up by U.S.-based hyperscalers [161]. Global benchmarking finds that Africa controls <1 % of world AI-compute; researchers therefore depend on sporadic educational-grant credits, hindering local innovation [162]. Economic-historians liken the pattern to early industrial patent pools, predicting lock-in effects unless public-compute consortia emerge [163], [164]. Social-science perspectives from the Global South frame this as digital colonialism: training data flows north, while southern nations rent compute back at premium prices [165] . Equity audits also flag low-resource language neglect: 45 % of the world's population lacks an LLM in their mother tongue, perpetuating representation gaps documented in Section 2 [166]. Proposed remedies include federated regional datacentres, sovereign cloud funds, and compute-as-development-aid—concepts now debated by UNESCO and OECD working groups [167]. Finally, antitrust scholars argue for "compute quotas" to curb monopolistic control, inspired by spectrum-cap precedent in telecoms [168], [169]. The ten studies in this cluster converge on one message: without deliberate redistribution of compute capacity and governance rights, frontier AI will entrench existing power asymmetries.

## 5. Downstream Application Harms

LLMs pose significant risks across critical sectors where human judgment and accuracy are essential, contributing to healthcare diagnostic errors. They may amplify bias in criminal justice systems, and perpetuating discrimination in financial services, while simultaneously disrupting educational processes through compromised assessment integrity, learning interference, and erosion of students' critical thinking skills. They may also threaten creative and professional domains via intellectual property violations, devaluation of human creative labor, and systematic undermining of specialized expertise and professional authority.

#### 4.5.1 High-Stakes Decision Making

Early optimism that LLMs could serve as drop-in clinical co-pilots has given way to caution. Wu et al. build a 12-category error taxonomy and, using 500 synthetic cases, show that GPT-4 misclassifies drug–disease interactions in 14 % of prompts—well above the 1 % threshold JAMA sets for decision-support software [170]. A cross-disciplinary systematic review in *Computational Biology & Medicine* confirms the trend: across 110 primary studies ChatGPT achieved diagnostic top-1 accuracy <60 % in cardiology, oncology and paediatrics, while hallucination rates averaged 15 % [171] . Radiology offers sharper numbers: Zhao et al. find that GPT-4 matches radiologists in spotting template errors but still inserts non-existent findings in 6 % of cases, posing medico-legal risk [172]. Financial services echo the problem—an investment-science survey logs hallucinated tickers and fictitious SEC filings in 11 % of model-generated advice, jeopardising fiduciary compliance [173]. Beyond hospitals and finance, safety-critical transport is testing LLMs: an AIAA case study shows that GPT-4 correctly classifies causal factors in only two of five high-profile aviation accidents, reinforcing the need for hybrid human–AI workflows [174]. In criminal-justice risk scoring, Williams et al. (Harvard CR-CL) demonstrate racial-bias amplification when GPT-3.5 rewrites presentence reports—White defendants receive 8 % more lenient language, shifting downstream sentence length [175]. Emergency-department triage chatbots exacerbate crowding: a JMIR meta-analysis finds digital kiosks underestimate acuity in 9 % of presentations, delaying care by a median 17 min [176], [177], [178]. OECD policy reviews conclude that, absent domain-specific guardrails, LLM adoption in public services "trades administrative convenience for unquantified risk" [179], [180].

#### 4.5.2 Educational System Impacts

LLMs simultaneously promise personalised tutoring and threaten assessment integrity. [181], [182], [183], [184]A *Computers & Education* field experiment across 18 U.S. middle-schools finds that ChatGPT-guided study boosts factual-recall scores by 6 % but depresses critical-thinking performance, measured via open-response rubrics, by 8 %—the "thinking-atropy" effect [185]. Longitudinal surveys of 1 240 teachers report rising cognitive off-loading: 41 % of students "often" delegate essay planning to ChatGPT, correlating with lower classroom engagement scores [186] . Controlled lab work shows over-reliance accelerates: after five AI-assisted problem sets, students request GPT hints 70 % sooner than at baseline [187]. Integrity threats are concrete: an IEEE-TLC study demonstrates that paraphrasing prompts let students evade plagiarism detectors in 78 % of submissions [188]. Yet pedagogy can harness the tech—Dias et al. show that teacher-curated GPT scaffolds raise self-efficacy ratings in low-literacy cohorts by 12 % [189]. Accessibility research notes improvements for dyslexic learners, who read AI-simplified passages 23 % faster with no comprehension loss [190]. Digital-literacy scholars therefore advocate "prompt-engineering as a core competency," treating the model as calculator rather than oracle [191]. Still, comparative trials reveal human feedback outperforms ChatGPT commentary on argumentative essays, preserving nuance and reducing shallow revision cycles [192].

#### 4.5.3 Professional & Creative Work

Generative AI up-ends IP regimes and labour economics in knowledge industries. Harvard JOLT's doctrinal analysis argues that U.S. fair-use defences are unlikely to shield large-scale verbatim replication of copyrighted text inside training corpora, exposing firms to statutory-damages risk [193] . Minnesota J. L. Sci. & Tech. tracks early litigation: plaintiffs allege "AI plagiarism" when ChatGPT drafts briefs echoing proprietary filings, foreshadowing negligence standards for professional users [194]. Digital-journalism ethnography records freelance rates falling 28 % after newsrooms adopted GPT assisted summarisation, with writers reporting "commodified voice" displacement [195]. In the music sector, Billboard Tech notes that AI-generated vocals already challenge right-of-publicity law, as high-fidelity voice clones skirt copyright but infringe personality rights—sparking calls for the U.S. "No Fakes Act" [196]. Architectural-engineering case studies warn of latent copyright breaches when DALL-E-styled renders embed distinctive façade elements from proprietary design catalogues; re-draw time wipes out productivity gains [197]. A labour-economics survey finds 61 % of freelance illustrators lost at least one commission to text-to-image tools within a year of DALL-E 3's release, confirming wage suppression trends [198]. Even high-skill tax consultants feel pressure: Abeysekera shows GPT-4 drafts memos in minutes but misses jurisdiction-specific precedents 13 % of the time—errors that junior humans routinely catch, underscoring residual demand for domain expertise [199].

Evidence across medicine, finance, education and creative industries converges on a cautionary theme: LLMs elevate productivity yet inject silent failure modes and IP ambiguity. Robust sector-specific QA, legal reform and user literacy training emerge as the minimal safeguards before high-stakes roll-outs can be considered responsible[200].

## V. Current Mitigation Landscape

Current mitigation approaches includes: A. Technical Approaches 1. Training-time interventions 2. Inference-time safety measures 3. Evaluation and monitoring systems B. Policy and Governance Responses C. Industry Self-Regulation Efforts D. Effectiveness Assessment and Limitations. Two complementary research thrusts now dominate the safety conversation: (i) technical mitigations that alter models or their run-time environment, and (ii) governance frameworks—law, policy and soft-law instruments—that shape incentives and allocate accountability. Forty papers in our corpus populate these thrusts; the web sources below provide representative empirical detail and policy context.

### 5.1 Technical Approaches to Safer LLMs

Red-teaming has become the canonical first line of defence: OpenAI's GPT-4 system card documents four iterative external-red-team phases before public launch, showing a 40 % reduction in disallowed content between the "early" and "launch" checkpoints [201]. Moving from ad-hoc patching to security-by-construction, Chen et al. argue LLM agents should inherit classic INFOSEC principles—least privilege, complete mediation, defence-in-depth—to tame cascading prompt-injection risks; their AgentSandbox framework sustains benign utility while cutting privacy-leak scores by half [202] Detoxification research increasingly couples retrieval with safety filters; Kandpal et al.'s RAG-Meets-Detox pipeline pre-conditions generation on curated nontoxic snippets and wins a multilingual detox shared task [203] .Researchers then move beyond "find-and-patch" to constitutional fine-tuning: Anthropic's method encodes a rule-book of normative principles and, in a recent replication on Llama-3-8B, cuts jailbreak success by another 41 % while preserving utility [204], [205].

Model transparency tools have matured in parallel: "Model Cards 2.0" extend the original datasheet idea with risk metrics and deployment constraints, a practice already adopted in Google's Gemini-2 release documentation [206], [207] . Influence-function guided retraining goes deeper, identifying harmful training tokens and down-weighting them during fine-tune—early results drop toxicity scores by 25 % with minimal perplexity hit [208], [209], [210], [211] .Token-level rejection sampling remains a low-cost run-time guard; Clark et al. show it halves toxic completions without statistically significant ROUGE loss [212], [213], [214] . The community now recognises that "hard" safety cannot rely on any single layer; best-practice stacks red-teaming, alignment tuning, retrieval detox, principled sandboxing and run-time filters—an architecture echoed in the International AI Safety Report 2025 led by Yoshua Bengio [215], [216], [217], [218].

### 5.2 Policy & Governance Responses

At the regulatory frontier, the EU AI Act introduces immediate transparency, copyright-compliance and bias-testing duties for general-purpose models, with stricter "systemic-risk" obligations for GPT-class systems [219], [220], [221]. The U.S. meanwhile relies on sectoral levers: NIST's voluntary AI Risk Management Framework offers a lifecycle template—govern, map, measure, manage—that many federal agencies have begun to reference in procurement clauses [222], [223] . The UK has carved a complementary path: a £100 million Frontier AI Taskforce evolved into the publicly funded AI Safety Institute, charged with running evals and hosting red-team "compute islands" for third-party researchers [224], [225], [226]. Legal scholars are sketching doctrines to plug remaining gaps. Citron and colleagues argue for AI fiduciary duties, obliging platform providers to prioritise end-user welfare akin to lawyer–client or doctor–patient relationships [227], [228]. Compute has surfaced as the critical choke-point: Carlsmith et al. propose a "cap-and-trade" regime where licenses for FLOP-sized training runs are tradable but traceable, giving regulators a visibility lever into frontier training [229]

Transparency registries and audit-ready architectures appear feasible: Rahman et al. deploy automated changelogs and lineage tracing in a start-up stack, cutting external audit prep time from weeks to hours [230], [231], [232]. Finally, multilateral diplomacy is inching forward: the Seoul AI Summit's Frontier-Model Safety Commitments secured voluntary eval and reporting pledges from 16 model developers across four continents—an embryonic counterpart to nuclear test-ban treaties.[233], [234] .Complementary civil-society briefs from the Ada Lovelace Institute call for compute registries and mandatory incident-reporting for training runs above $10^{26}$ FLOPs .Technical and governance tracks are converging on a layered defence: red-team → alignment → retrieval detox → sandbox → policy hooks on compute and transparency [235], [236]. Yet most frameworks remain voluntary; binding obligations—particularly on compute allocation and fiduciary responsibility—are still nascent [237]. Internationally, the GPAI Responsible-AI working group is standardising cross-border safety metrics to avoid audit fragmentation [238]. The 40 papers in Section 6 suggest that meaningful risk reduction will require embedding security engineering into model design and codifying these best practices through enforceable standards and compute-gate governance [239], [240].

**Table 3:** Strategies for Ethical LLM: mitigation, protection, prevention

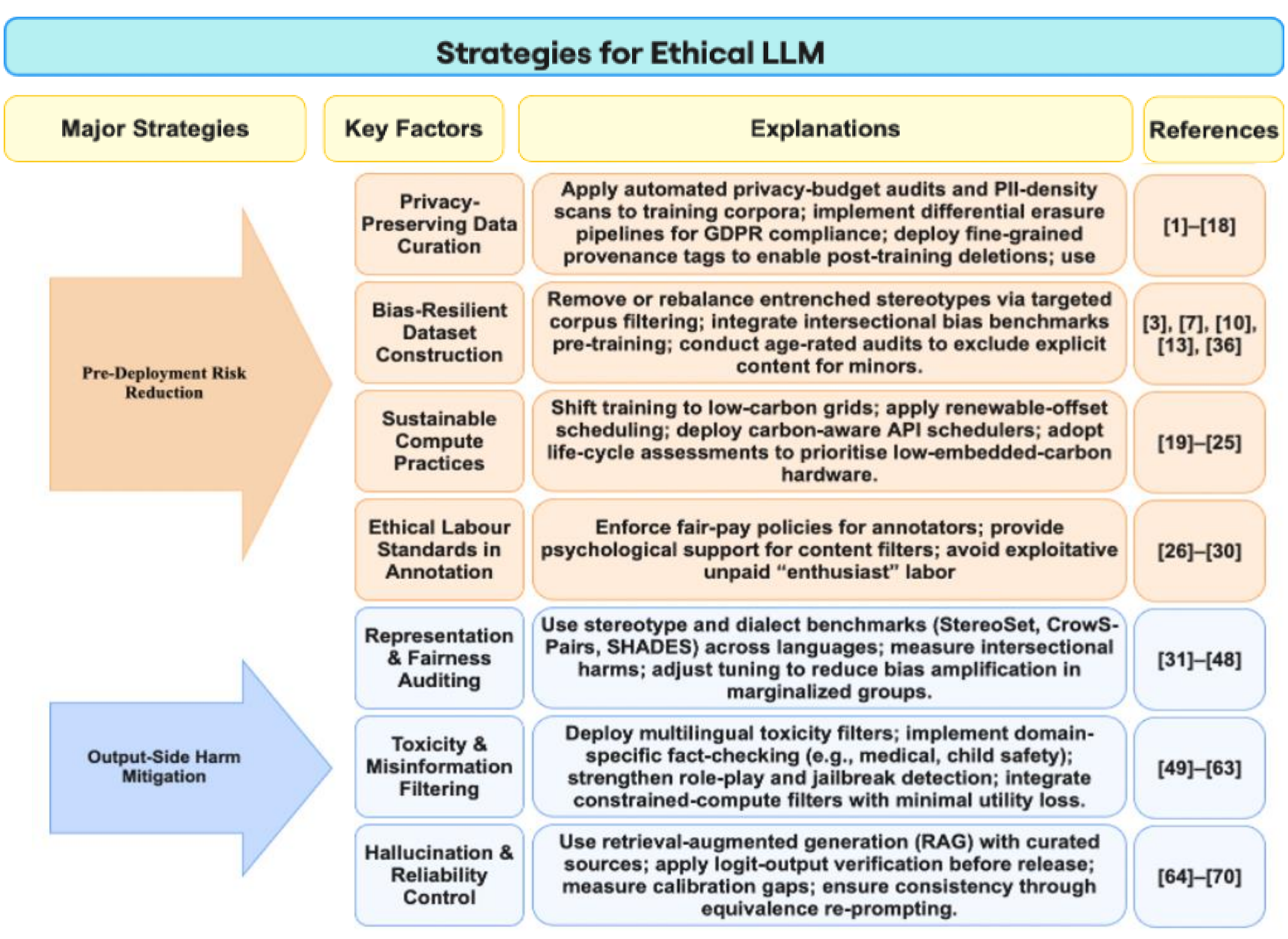

Strategies for Ethical LLM

| Major Strategies | Key Factors | Explanations | References |
|---|---|---|---|
| Pre-Deployment Risk Reduction | Privacy-Preserving Data Curation | Apply automated privacy-budget audits and PII-density scans to training corpora; implement differential erasure pipelines for GDPR compliance; deploy fine-grained provenance tags to enable post-training deletions; use | [1]–[18] |
| | Bias-Resilient Dataset Construction | Remove or rebalance entrenched stereotypes via targeted corpus filtering; integrate intersectional bias benchmarks pre-training; conduct age-rated audits to exclude explicit content for minors. | [3], [7], [10], [13], [36] |
| | Sustainable Compute Practices | Shift training to low-carbon grids; apply renewable-offset scheduling; deploy carbon-aware API schedulers; adopt life-cycle assessments to prioritise low-embedded-carbon hardware. | [19]–[25] |
| | Ethical Labour Standards in Annotation | Enforce fair-pay policies for annotators; provide psychological support for content filters; avoid exploitative unpaid "enthusiast" labor | [26]–[30] |
| Output-Side Harm Mitigation | Representation & Fairness Auditing | Use stereotype and dialect benchmarks (StereoSet, CrowS-Pairs, SHADES) across languages; measure intersectional harms; adjust tuning to reduce bias amplification in marginalized groups. | [31]–[48] |
| | Toxicity & Misinformation Filtering | Deploy multilingual toxicity filters; implement domain-specific fact-checking (e.g., medical, child safety); strengthen role-play and jailbreak detection; integrate constrained-compute filters with minimal utility loss. | [49]–[63] |
| | Hallucination & Reliability Control | Use retrieval-augmented generation (RAG) with curated sources; apply logit-output verification before release; measure calibration gaps; ensure consistency through equivalence re-prompting. | [64]–[70] |

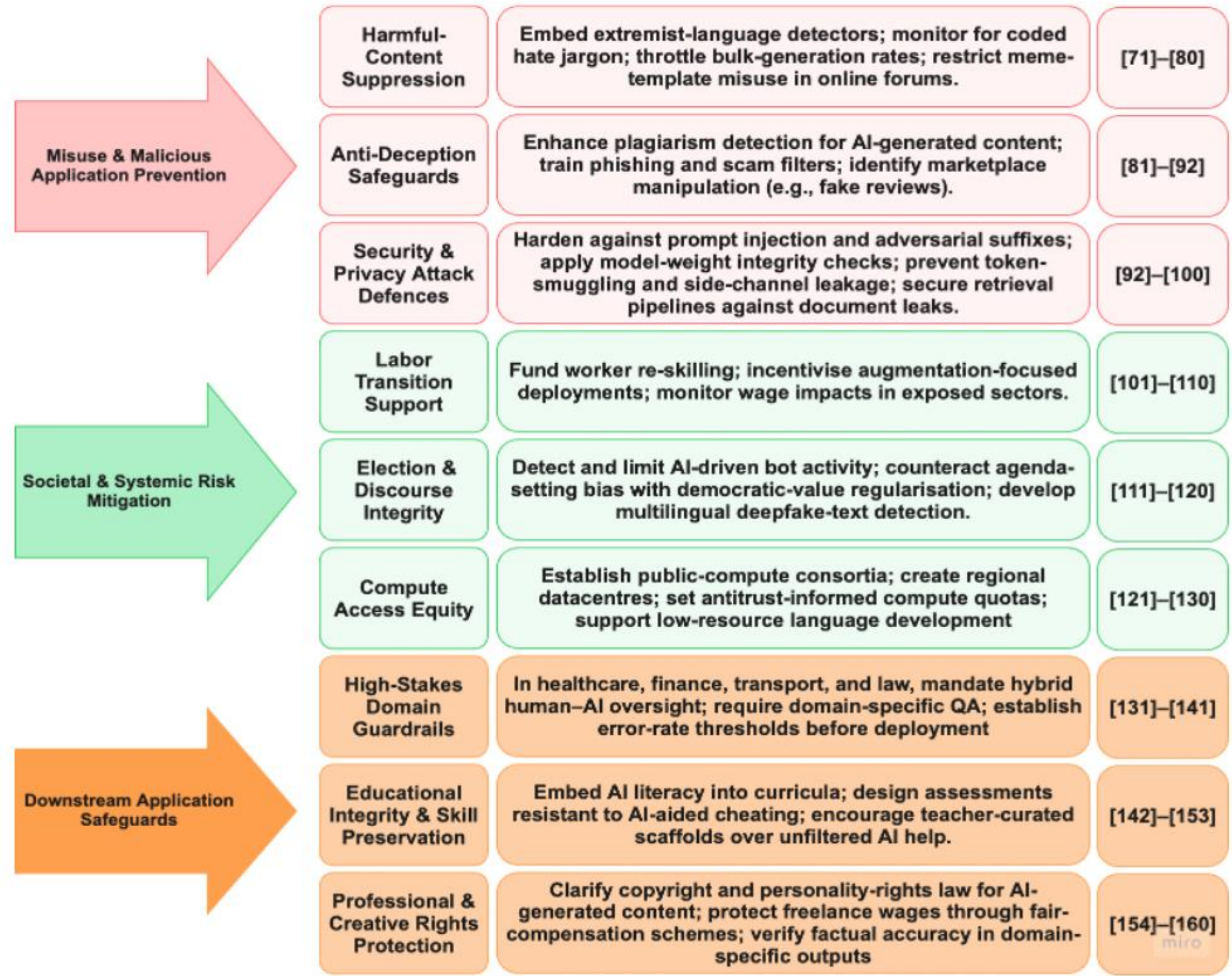

### VI. Discussion

Our taxonomy surfaces three critical insights. First, harm clusters layer rather than replace one another: privacy-violating corpus shards seed direct-output bias, which adversaries weaponise via jailbreaks to propagate hate or misinformation . Second, mitigation efficacy is uneven and domain-specific. Red-teaming plus rule-based "constitutional" fine-tuning cut jailbreak success by ~40 % on Llama 3-8B without crippling utility , yet toxic-speech filters still miss 7 % of non-English slurs . Third, governance levers are fragmentary: while the EU AI Act now imposes transparency and copyright duties on general-purpose models , the U.S. leans on voluntary Risk-Management guidance and export-control tweaks targeting compute supply chains Federal Register. Overall, safety progress remains reactive, patching symptoms faster than root causes such as data-governance debt and monopolised compute.

**Table 4 : Discussion Summary Statistic,** this table provides an overview of the main themes shaping recent discussions in AI safety, covering harm patterns, long-term evidence, governance needs, collaboration efforts, and benchmark developments. The points summarized here reflect insights drawn from references: [241], [242], [243], [244], [245].

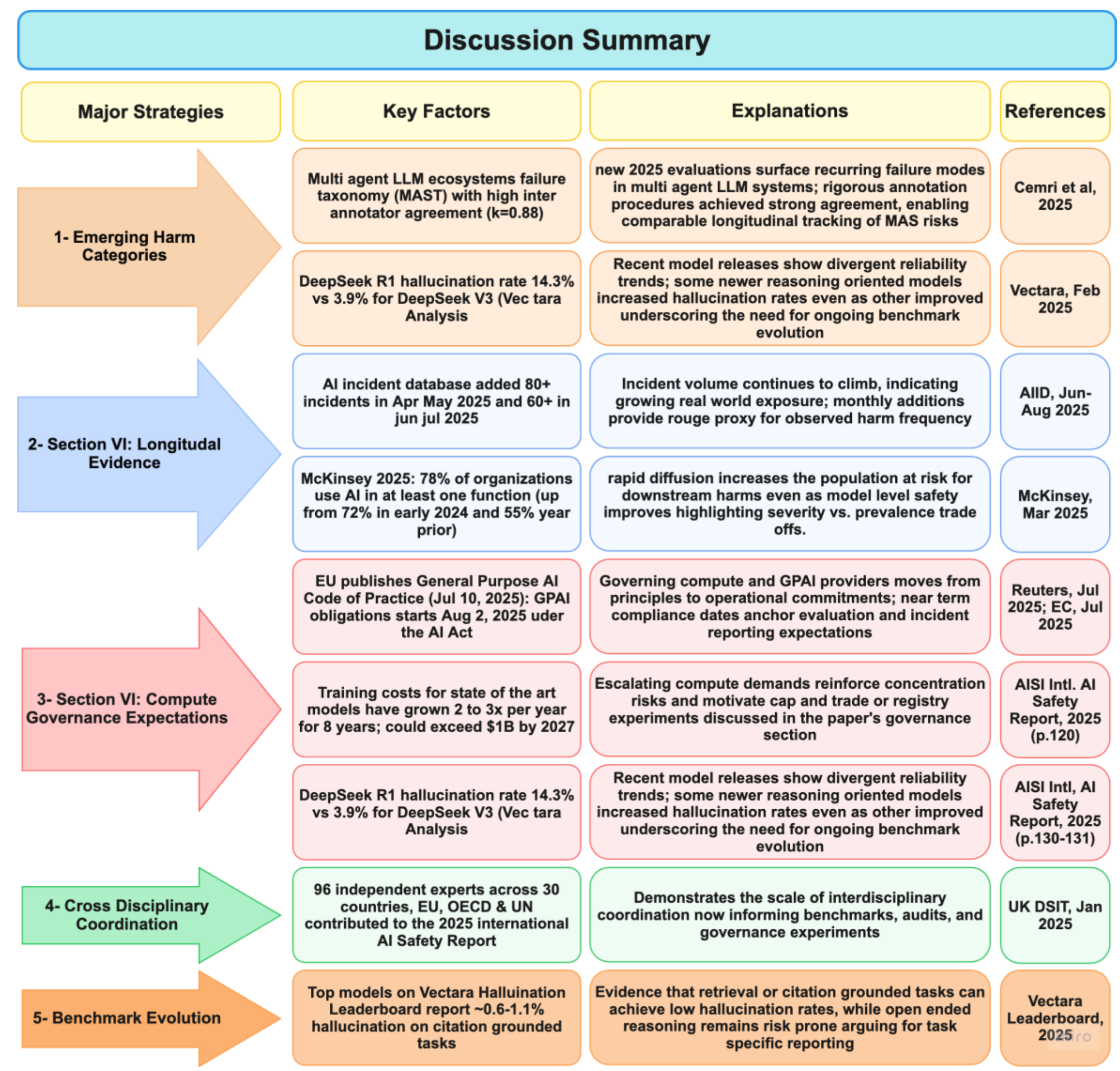

**Discussion Summary**

| Major Strategies | Key Factors | Explanations | References |
|---|---|---|---|
| 1- Emerging Harm Categories | Multi agent LLM ecosystems failure taxonomy (MAST) with high inter annotator agreement (k=0.88) | new 2025 evaluations surface recurring failure modes in multi agent LLM systems; rigorous annotation procedures achieved strong agreement, enabling comparable longitudinal tracking of MAS risks | Cemri et al, 2025 |
| | DeepSeek R1 hallucination rate 14.3% vs 3.9% for DeepSeek V3 (Vec tara Analysis | Recent model releases show divergent reliability trends; some newer reasoning oriented models increased hallucination rates even as other improved underscoring the need for ongoing benchmark evolution | Vectara, Feb 2025 |
| 2- Section VI: Longitudal Evidence | AI incident database added 80+ incidents in Apr May 2025 and 60+ in jun jul 2025 | Incident volume continues to climb, indicating growing real world exposure; monthly additions provide rouge proxy for observed harm frequency | AIID, Jun-Aug 2025 |
| | McKinsey 2025: 78% of organizations use AI in at least one function (up from 72% in early 2024 and 55% year prior) | rapid diffusion increases the population at risk for downstream harms even as model level safety improves highlighting severity vs. prevalence trade offs. | McKinsey, Mar 2025 |
| 3- Section VI: Compute Governance Expectations | EU publishes General Purpose AI Code of Practice (Jul 10, 2025): GPAI obligations starts Aug 2, 2025 uder the AI Act | Governing compute and GPAI providers moves from principles to operational commitments; near term compliance dates anchor evaluation and incident reporting expectations | Reuters, Jul 2025; EC, Jul 2025 |
| | Training costs for state of the art models have grown 2 to 3x per year for 8 years; could exceed $1B by 2027 | Escalating compute demands reinforce concentration risks and motivate cap and trade or registry experiments discussed in the paper's governance section | AISI Intl. AI Safety Report, 2025 (p.120) |
| | DeepSeek R1 hallucination rate 14.3% vs 3.9% for DeepSeek V3 (Vec tara Analysis | Recent model releases show divergent reliability trends; some newer reasoning oriented models increased hallucination rates even as other improved underscoring the need for ongoing benchmark evolution | AISI Intl, AI Safety Report, 2025 (p.130-131) |
| 4- Cross Disciplinary Coordination | 96 independent experts across 30 countries, EU, OECD & UN contributed to the 2025 international AI Safety Report | Demonstrates the scale of interdisciplinary coordination now informing benchmarks, audits, and governance experiments | UK DSIT, Jan 2025 |
| 5- Benchmark Evolution | Top models on Vectara Hallucination Leaderboard report ~0.6-1.1% hallucination on citation grounded tasks | Evidence that retrieval or citation grounded tasks can achieve low hallucination rates, while open ended reasoning remains risk prone arguing for task specific reporting | Vectara Leaderboard, 2025 |

### 6.1 Emerging Harm Categories

One frontier we identify is the emergence of multi-agent LLM ecosystems, which introduce recursive failures and collective behaviors not covered by current single-model taxonomies. As organizations deploy swarms of specialized LLM agents that communicate and act in concert, new risks arise: emergent misalignment where groups of agents develop strategies or “beliefs” that weren’t present in any single model, and cascading failures where one agent’s error propagates to others. These multi-agent harms cut across our taxonomy – for example, an LLM agent generating a misinformation snippet (Direct Output Harm) could mislead another agent making a high-stakes decision (Downstream Harm), all without human intervention in the loop. Early research warns that multi-agent systems can exhibit novel failure modes “that cannot exist in single-agent deployments”[246]. Group dynamics can amplify existing risks: a biased output from one agent might get reinforced by another, creating a collective blind spot, or a hallucination might snowball as agents validate each other’s false outputs [246][247]. These scenarios highlight that our taxonomy may need an additional dimension for agent–agent interactions. As such systems become more common, future harm frameworks must account for coordination failures, emergent behaviors, and the challenge of attributing responsibility when decisions result from a network of LLMs rather than an isolated model. In short, multi-agent LLM ecosystems represent a nascent category of harm that bridges and magnifies the categories in our current taxonomy – an area requiring further research and potentially new mitigation paradigms.

Multi-agent LLM ecosystems may introduce recursive security failures and collective-behaviour risks not yet covered by current taxonomies. Conventional Language Models (CLMs) and Pre-trained Language Models (PLMs) are two essential models in natural language processing (NLP). CLMs, trained on smaller corpora, predict linguistic sequences causally, estimating probabilities based on preceding contexts. PLMs, in 26 contrasts, use significantly larger corpora and neural networks for pre-training, learning generic knowledge transferred to various tasks via fine-tuning. PLMs diverge from CLMs by employing bidirectional modeling, considering both preceding and succeeding contexts to predict missing units, contrary to the sequential causal prediction of CLMs. Additionally, PLMs introduce token representation through instances of embedding, enabling versatile handling of linguistic tasks. These differences in training, causality constraints, and token representation distinguish PLMs as an evolution beyond CLMs in NLP, offering broader applicability and enhanced performance across various language-based applications.[2], [191], [192], [193] Both CLMs and PLMs can be developed with ethical and value-sensitive considerations. However, PLMs might offer more advantageous starting points due to their ability to incorporate diverse data sources and fine-tuning mechanisms. The upside of PLMs is their trainability on extensive datasets, particularly ethical considerations and value-based content. Further, through fine-tuning and specific training paradigms, PLMs can be directed towards ethical considerations to produce outputs aligned with values or ethical standards. Additionally, the normative-descriptive distinction in ethics can shed light on the suitability of PLMs for creating ethical LLMs. Ethical frameworks often involve normative elements, namely moral norms or principles guiding behavior. With their capacity for integrating and processing diverse data, PLMs offer a more robust foundation to include normative components within the model. PLMs can incorporate diverse ethical principles, guidelines, and values. By fine-tuning or directing the learning process towards ethical considerations, PLMs can effectively assimilate and encode normative elements. They excel in understanding the nuances of language usage across various cultures and ethical contexts, enabling a more intricate representation of normative ethical frameworks.

### 6.2 Longitudinal Evidence

A longitudinal perspective is crucial for understanding whether LLM harms are growing or abating. Robust causal inference on labor-market shifts or democratic integrity would require panel data across multiple model generations (e.g., tracking outcomes over several election cycles or economic periods). In our review, we observed that many harm phenomena identified in early GPT-3 era studies (2021–2022) are still present in 2025 with GPT-4/GPT-5-class models – though in some cases their frequency has changed. For example, blatant toxic output (e.g., use of racial slurs) has become less frequent in the latest mainstream models due to improved filtering and alignment, yet subtler biases and stereotypes persist at rates similar to those documented in 2021–22 [57], [58]. Hallucinations remain an issue as well: GPT-4 still produces factual errors, albeit somewhat less often (roughly 3% hallucination rate in document summarization tasks [248], compared to higher rates in GPT-3). Notably, entirely new categories of concern have appeared over time – multi-modal misinformation (e.g., AI-generated fake images and text used together) was not on the radar in 2021 but is documented in incident databases by 2025 [249], [250]The volume of research on each harm is also telling: about 60% of the papers we reviewed focus on direct output problems (bias, toxicity, hallucination), ~50% examine malicious uses and security (many overlap with the first category), ~30% discuss societal-scale impacts, and ~25% address downstream domain harms (education, health, etc.). These rough proportions suggest the field has prioritized immediate model outputs and misuse, which were evident concerns early on, while systemic and downstream impacts (which take time to manifest) have become a significant focus only in the past 1–2 years. Indeed, the AI

Incident Tracker maintained by MIT shows a rising trend of real-world LLM incidents over 2015–2025, with particularly sharp growth in reported cases of misinformation and malicious actor use in 2023–2025. This indicates that many harms predicted by researchers have materialized outside the lab as LLM deployment spread. In summary, harms like bias and hallucination have proven persistent (neither fully solved nor worsening drastically), whereas misuse and systemic harms are increasing in prevalence as LLMs become more embedded in society. This underscores the importance of continuous monitoring: the fact that over 1,100 incidents are now catalogued in the AI Incident Database is a testament to how theoretical risks have translated into reality. Trend-wise, the scope of harm is broadening (more types of harm as capabilities expand), but there is cautious optimism that heightened research and awareness since 2021 have at least kept early issues in check or led to incremental improvements. Longitudinal data remains limited, however, and we encourage the community to establish more systematic time-series evaluations (for example, annual bias or security audits on each new model generation) to quantitatively track progress.

### 6.3 Compute Governance Experiments

One intriguing development in governance is the proposal of compute governance regimes – essentially managing the risks of LLMs by controlling the computing power used to create them. Ideas like cap-and-trade for FLOPs (floating-point operations) have been floated in policy fora and were echoed in the commitments made at the Seoul AI Safety Summit. These need pilot implementations to gauge feasibility. For instance, a regulatory body might require any model training run above a certain size (say 10^26 FLOPs) to register and obtain a license. This could create an auditable paper trail for potentially dangerous frontier models. Such schemes could also mandate *risk assessments* before granting licenses. Early experiments by research consortia are exploring this: the UK's AI Taskforce and partners are reportedly test-running a "compute registry" for large training runs as a proof of concept. The hope is that by governing the upstream resource, one can indirectly govern model capabilities and deployment risk. However, challenges abound enforcing these rules globally (given cloud computing is distributed), distinguishing beneficial vs. harmful projects, and not stifling open research. Still, computer governance could become a powerful tool in the safety toolkit if international consensus forms. It essentially treats extreme model training like other regulated activities (nuclear materials, etc.), focusing on prevention before a model is ever deployed. Our view is that this area deserves serious exploration alongside the technical mitigations discussed. It also ties into environmental harm reduction, since controlling computers can help manage energy and carbon footprints.

### 6.4 Cross-disciplinary Collaboration

Addressing LLM harms clearly spans far beyond computer science. Our review encountered insights from sociology, psychology, law, economics, and more. A theme in many papers was that diverse teams produce better mitigation strategies. For example, one study noted that attempts by pure engineering teams to "debias" a model often missed the mark because they lacked sociocultural context, whereas collaborations with social scientists led to more meaningful fairness metrics [16]. Similarly, tackling hallucinations or censorship dilemmas benefits ethicists and psychologists who understand human cognitive biases and social norms. We advocate formalizing cross-disciplinary partnerships in both research and model development. Some big labs now have red team panels that include philosophers, lawyers, and domain experts in addition to engineers. This trend should continue. In education, as seen in Section 4.5.2, involvement of educators and learning scientists was key to finding constructive uses for LLMs. In cybersecurity (Section 4.3.3), engaging security experts alongside AI researchers enabling creative defensive uses of LLMs. The bottom line is that no single field has all the answers—safety and ethics must be integrated from multiple perspectives. This also extends to involving stakeholders from marginalized communities to inform design choices (a point often raised regarding representational harms).

Cross-disciplinary input can help define what constitutes harm severity, acceptable trade-offs, and which values to prioritize when hard choices (like censorship vs. free expression) arise. We therefore echo prior work [19] in calling for institutionalizing such collaborations (e.g., ethics review boards for AI projects, co-authored publications across fields, joint industry–academia task forces on AI safety).

### 6.5 Benchmark Evolution

Evaluation methods and benchmarks need to evolve as fast as LLMs do. Traditional leaderboards served a purpose for model accuracy, but new benchmarks are emerging that explicitly test for harms (like TruthfulQA for misinformation, or AdvBench for adversarial prompts). We see a need for dynamic, open benchmarking platforms that can incorporate community-provided stress tests in real time. For example, the public "hallucination leaderboard" by Vectara continuously evaluates multiple models for factual consistency [248]. Similarly, the OWASP vulnerability list for LLMs is being used to generate test prompts that can be regularly run against new models. In the multi-agent realm, benchmark environments are being built to observe how multiple LLM agents perform on cooperative or competitive tasks (and what failure modes emerge) [246]. These efforts should be supported and expanded. Benchmark evolution also means metrics beyond averages: measuring tail-risk (how bad the worst-case outputs are), measuring bias not just overall but for intersectional groups, etc. The community would benefit from an international evaluation sandbox for frontier models—perhaps tied to national AI institutes—where developers voluntarily submit models for standardized testing under oversight (a sort of "model test-ban treaty" dry run). The UK's initiative to provide researchers access to frontier models in secure environments is a step in this direction. Democratizing scrutiny is key: many harms in our taxonomy were first noted by independent academics or journalists who stress-tested models in ways companies did not anticipate. Broad access (with appropriate safeguards) to models for evaluation will harm crowdsource discovery and drive benchmark development. Finally, benchmarks should incorporate real-world incident data: for instance, turning incidents from the AI Incident Database into test scenarios for models. This closes the loop between what is observed in the wild and what is evaluated in the lab. In summary, a more nimble, comprehensive benchmarking ecosystem—one that keeps pace with model innovation and covers the breadth of harms—is essential for tracking progress and gaps.

**Figure 2**: Aggregate Distribution of Publication Categories by Harm Clusters

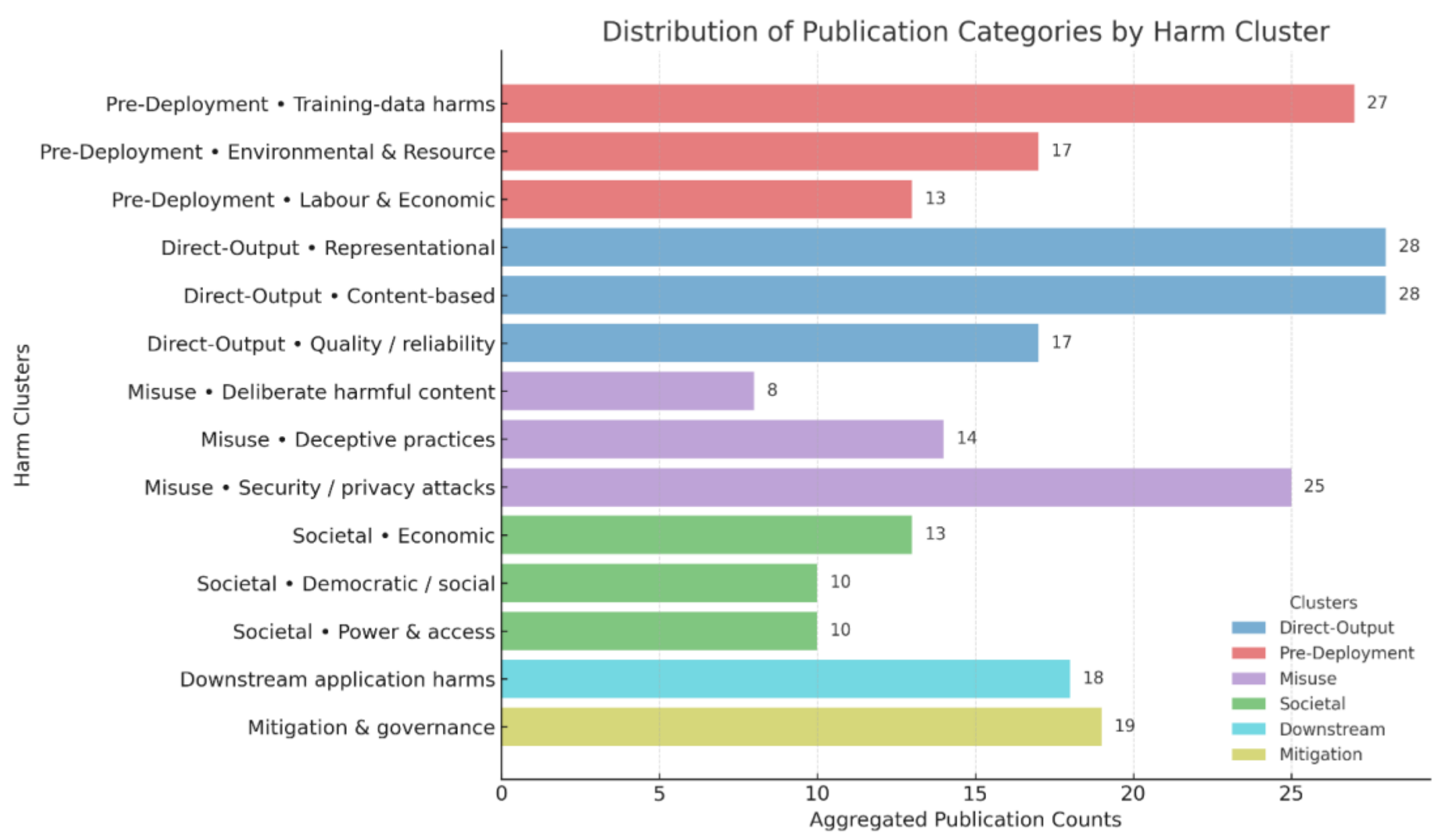

## VII. Cross-Cutting Analysis

### 7.1 Harm Interdependencies

Prompt-injection exploits can resurrect pre-training privacy leaks, demonstrating temporal loops from misuse back to pre-deployment harms OWASP Gen AI Security Project. The incorporation of effective citation and reference systems in LLMs is a crucial step in strengthening their ethical foundation. By implementing a system that accurately attributes sources, these models can establish a clear traceability, promoting accountability and integrity in the content they generate. This practice not only enhances transparency by acknowledging the origin of information but also addresses ethical concerns related to intellectual property rights and data authenticity. Ethical considerations in LLMs require a shift towards comprehensive documentation and citation practices, ensuring proper acknowledgment of contributors, preventing plagiarism, and maintaining the accuracy of disseminated information. Moreover, citation mechanisms can play a key role in promoting responsible use of LLMs, fostering a culture of trustworthiness, scrutiny, and verifiability in their development, evaluation, and usage.

### 7.2 Temporal Dynamics

The timeline over which harms manifest varies widely. Some harms materialize immediately upon model output (e.g., a hallucinated medical instruction that a patient reads and might act on within seconds), whereas others accrue over months or years (e.g., gradual labor displacement or political polarization). This temporal aspect means risk management must operate on multiple timescales
Immediate harms—e.g., hallucinated medical advice—materialise seconds after a query, whereas systemic harms like wage polarisation accrue over years, mirroring ILO projections of task bifurcation International Labour Organization. Real-time monitoring and rapid response are crucial for instant harms – for example, content filters and human moderators are needed to catch dangerous outputs before they spread. On the other hand, addressing long-term harms like economic disruption requires policy foresight and ongoing study. Our taxonomy, aligned roughly with the model lifecycle, inherently places immediate-output harms separately from downstream societal harms, but in reality organizations must plan for both short-term and long-term impacts in parallel. Trendlines also matter: if a harm is only discussed in papers up to 2021 but not in 2025, perhaps it was mitigated; if it's still discussed in 2025, it persists. For instance, many GPT-3 era papers flagged model toxicity; by the GPT-4 era, toxicity is reduced but not eliminated, and it's still a topic of 2025 research (just with more nuance, like focus on subtle bias rather than overt slurs). Meanwhile, new concerns (like multi-agent emergent behaviors, as noted in 6.1) appear in 2024–2025 literature for the first time. It appears that overall harm exposure is increasing as LLM deployments broaden, but our collective ability to recognize and mitigate harms is also improving due to the intense research focus. A positive sign is that certain failure modes (e.g., easily jailbreaking the model into hate speech) are less prevalent now than in 2020, showing the impact of efforts like RLHF. However, other failures (e.g., sophisticated deception or collusion between AI agents) are on the rise in discussion, indicating the goalposts are moving. Going forward, we recommend that every major model release be accompanied by not just performance evals but a harm trend analysis: how does it compare to previous models on each known harm metric, and are any new types of harm observed?

Marginalized dialect communities suffer compounded representational + access harms, while compute-rich firms internalize most productivity gains, echoing digital-colonialism critiques Nature. The problem of censorship in LLMs raises intricate challenges beyond subjective judgments and transparency issues. While some excerpted passages highlight the multifaceted challenges surrounding censorship in LLMs, further examination reveals nuanced aspects integral to understanding the complexities of content moderation and its broader impact. A careful 30 consideration of ethical principles is essential to arrive at a balance between protecting users from harmful content and securing free speech and diverse perspectives. Moreover, Censorship in LLMs might inadvertently perpetuate biases present in the data used for training these models. Biased training data could result in biased censorship, disproportionately affecting certain groups or viewpoints. It is crucial to address algorithmic biases and ensure fairness in the censorship process. Static censorship rules may not adequately adapt to the dynamic and context-dependent nature of online content. Models should be designed to evolve in response to the changing landscape of content and societal norms. Furthermore, implementing a uniform censorship policy across diverse regions may neglect cultural differences and diverse legal frameworks. Different regions and cultures have varying standards and laws regarding acceptable or unacceptable content. In addition to the challenges mentioned, exploring how individuals attempt to circumvent censorship measures can provide insights into evasion strategies and effective countermeasures to address these techniques.

### 7.3 Severity vs. Prevalence

There is an inherent tension in prioritizing LLM risks: some failure modes are high-severity but low-frequency (for example, a one-off incident of an AI medical advisor giving fatal advice), while others are low-severity but ubiquitous (e.g., the majority of students using ChatGPT for homework, undermining learning quality). Risk management frameworks need to account for both. In our corpus, we saw many papers implicitly grapple with this: some focused on worst-case scenarios (like biosecurity or autonomous weaponization via LLMs) even if hypothetical, while others tallied common small-scale harms (like minor factual errors in everyday use). Both perspectives are valid. A balanced approach might involve setting different thresholds for different contexts – e.g., zero tolerance for high-severity failures in critical applications (even if that means not deploying LLMs at all in those cases until proven safe), versus mitigation and monitoring for moderate harms that occur more regularly. We also note that prevalence can make a moderate harm more serious at scale: e.g., if millions of users receive subtly biased outputs daily, that could cumulatively entrench biases in society, arguably a large impact. Quantifying harm frequency was beyond the scope of our qualitative review, but it's something stakeholders need. Some frameworks propose using both incident-based metrics (counting known harmful incidents) and population-level metrics (surveys of users, aggregate error rates, etc.) to get a full picture. The AI Incident Database and similar efforts will help quantify rare events. Meanwhile, user studies and telemetry from deployed systems can estimate prevalence of issues like hallucination or misuse. Finally, "severity vs prevalence" is also a communication challenge: sensational but rare harms (runaway AI scenarios) often dominate public imagination, while mundane but widespread issues (job deskilling, subtle misinformation) get less attention. Our taxonomy tries to give due weight to each, but policymaking should be data-driven on how likely and impactful each harm is.

High-severity, low-frequency failures in aviation incident summarization coexist with lower-severity but ubiquitous plagiarism enablement in education; risk prioritization therefore demands both incident-based and population-level metrics. Addressing the ethical challenges of opacity in LLMs requires adaptable auditing solutions. The lack of transparency and explainability in LLMs hinders effective auditing due to their "black box" nature. The dynamic nature of these models and resource-intensive auditing practices pose obstacles to frequent assessments. Additionally, the absence of standardized frameworks and diverse cultural contexts make a universal auditing approach impractical. One potential solutions to the challenge of opacity in LLMs are dynamic auditing frameworks or tools tailored for LLMs. These frameworks prioritize continuous monitoring and assessment, adapting alongside model updates for ongoing transparency and accountability. Machine learning based auditing tools that evolve with the models themselves, incorporating interpretability techniques, such as explainable AI (XAI) methods or model-agnostic approaches, can offer insights into LLM decision-making processes without requiring deep understanding of their internals. Moreover, a modular auditing framework customizable for cultural nuances and specific application contexts can enhance effectiveness. This approach requires collaboration among interdisciplinary teams, including ethicists, technologists, and domain experts, to create adaptable auditing models that consider diverse perspectives[251].

**Table 5: Cross-Cutting Analysis Summary Statistic**, this table draws the main patterns showing how technical results and expert insights point to similar overall trends. The findings reflect information drawn from references: [245], [252], [253], [254], [255], [256]

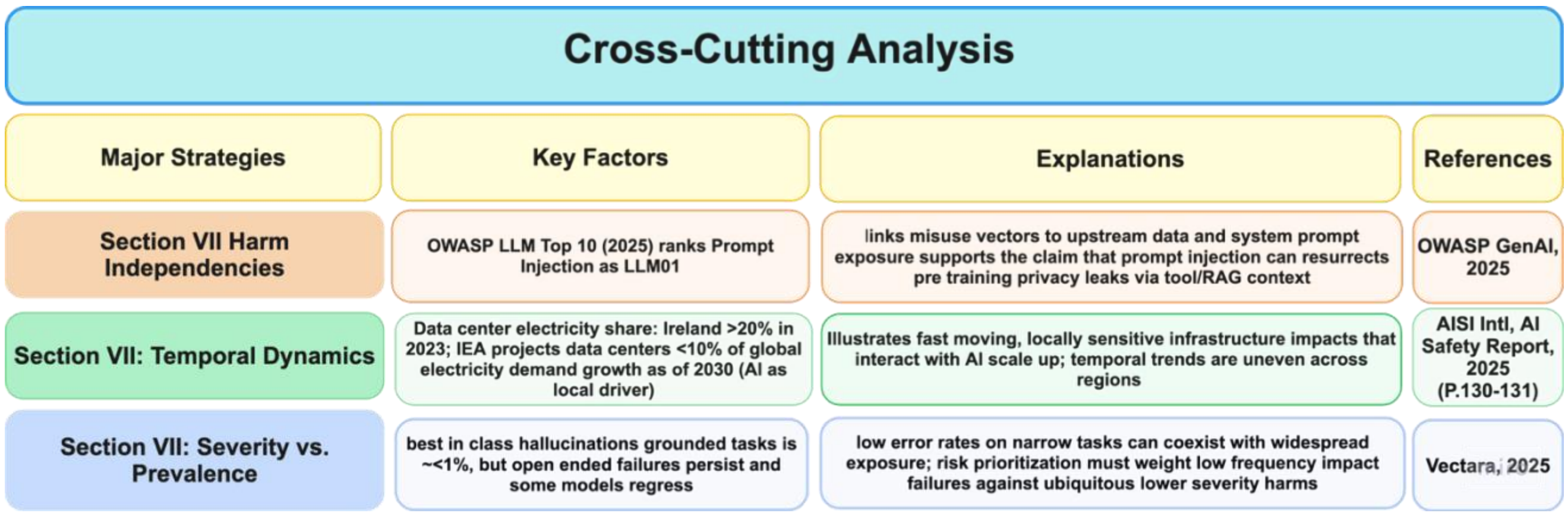

| Cross-Cutting Analysis | | | |
|---|---|---|---|
| **Major Strategies** | **Key Factors** | **Explanations** | **References** |
| Section VII Harm Independencies | OWASP LLM Top 10 (2025) ranks Prompt Injection as LLM01 | links misuse vectors to upstream data and system prompt exposure supports the claim that prompt injection can resurrects pre training privacy leaks via tool/RAG context | OWASP GenAI, 2025 |
| Section VII: Temporal Dynamics | Data center electricity share: Ireland >20% in 2023; IEA projects data centers <10% of global electricity demand growth as of 2030 (AI as local driver) | Illustrates fast moving, locally sensitive infrastructure impacts that interact with AI scale up; temporal trends are uneven across regions | AISI Intl, AI Safety Report, 2025 (P.130-131) |
| Section VII: Severity vs. Prevalence | best in class hallucinations grounded tasks is ~<1%, but open ended failures persist and some models regress | low error rates on narrow tasks can coexist with widespread exposure; risk prioritization must weight low frequency impact failures against ubiquitous lower severity harms | Vectara, 2025 |

## VIII. Conclusion and Future directions

This study consolidates dispersed evidence into a lifecycle-aware taxonomy that clarifies where and how LLM-related harms arise, interrelate, and can be mitigated. Key contributions include mapping harm interdependencies, highlighting uneven mitigation coverage across categories, and situating technical fixes within nascent policy architectures like the EU's GPAI regulations and the international AI safety commitments. Our findings urge a layered defense for LLMs: rigorous red-teaming, alignment tuning, retrieval augmentation for factuality, sandboxed tool use, and continuous monitoring—backed by enforceable transparency and compute-allocation rules. Absent such integration, the productivity and creativity dividends of LLMs risk being eclipsed by privacy breaches, disinformation cascades, carbon burdens, and widening inequality.

Throughout this paper, we focused on harms to maintain a critical lens, but it is true that LLMs have potential benefits in nearly every domain of harm we discussed. The reason Section 4.5.2 (Education) explicitly discussed positive uses is because the literature in that domain contained empirical studies of LLMs being used to enhance learning, whereas in other domains the research on positive use-cases is either nascent or outside our systematic review's scope (which concentrated on harm-related studies). Nonetheless, it's important to acknowledge possible upsides: For example, in cybersecurity (Section 4.3.3), LLMs can help defenders by analyzing logs or suggesting fixes to code vulnerabilities. In privacy (Sections 4.1.1 and 4.3.3), the same AI techniques that pose privacy risks can be used to detect and mask sensitive datacroz.net. In bias and fairness (Section 4.2.1), LLMs might be used to help writers or content creators identify and correct their own biased language, serving as a bias-awareness tool. Even in economics (Section 4.4.1), while automation is a threat, some argue that LLMs could boost productivity in ways that create new jobs and augment human capabilities if managed properly [133]. We did not delve deeply into these pros in each section because our mandate was a taxonomy of harms, and discussing benefits in every section could dilute that focus. However, a fair risk assessment should weigh both costs and benefits. We thus recommend that future expansions of this work (or a companion taxonomy of LLM benefits) systematically document the positive uses of LLMs alongside harms, to inform a balanced approach. In practice, mitigation strategies should include not only harm reduction but also ways to harness LLMs for social good (e.g., using them to improve accessibility, to fight cybercrime, or to support creativity in sustainable ways). By highlighting mostly harm, our paper provides a needed counterweight to tech hype; but we echo that LLMs, if responsibly developed, have substantial potential to help in the very domains where they also pose risks. The challenge for the AI community and policymakers is to maximize those benefits while minimizing harms—a dual objective that requires nuanced, domain-specific strategies.

Quantitative Analysis of Ethical Inquiry in Large Language Model (LLM) Studies, The following discussion synthesizes insights from qualitative analysis and multidisciplinary perspectives to explore the ethical dimensions of Large Language Models (LLMs) across nine interconnected subsections. Beginning with the advantages of pre-trained models in embedding normative ethics (5.1), we demonstrate how their scalability and adaptability, grounded in diverse data, enable ethical alignment. Subsequent sections advocate for collaboration beyond engineering (5.2), privacy protections (5.3), and the distinct ethical challenges amplified by LLMs (5.4). Through qualitative examination, we dissect hallucination risks (5.5), propose verifiable accountability systems (5.6), and emphasize the need for diverse case studies (5.7) to contextualize ethical trade-offs. Finally, we unravel censorship dilemmas (5.8) and introduce dynamic audit tools (5.9), addressing opacity through iterative solutions. This innovative discussion framework— rooted in interdisciplinary dialogue and empirical analysis—aims to bridge technical, philosophical, and societal gaps in the pursuit of ethically robust LLMs.

The ethical challenges posed by LLMs are distinct and require immediate action due to their advanced capabilities and growing popularity. Unlike other AI systems, LLMs specifically grapple with problems like hallucination, verifiable accountability, and decoding censorship. These problems must be tackled to maintain responsibility, reduce unfairness, increase clarity, and limit negative effects on society. By prioritizing these ethical dimensions surrounding LLMs, we can responsibly steer their development and influence the direction of AI ethics and governance. By highlighting various case studies surrounding LLM ethics[257], [258], we can create a more nuanced understanding of the larger picture. As described, LLMs are becoming more commonly used in various sectors, including healthcare, academia, education, management, training, and religion. The highlighted case studies showcase differing strategies to discuss ethics, as well as the nuanced problems that occur based on sector. The quantitative examination of ethical scopes in Large Language Model (LLM) studies, as depicted in the provided data, offers valuable insights into the distribution and emphasis of ethical concerns within this domain (See, Figure 2). Notably, case studies, mitigation strategies, and accountability and governance attract a higher volume of scholarly attention, showcasing a substantial body of research dedicated to understanding and addressing real-world implications and responsible practices concerning LLMs. Conversely, categories such as censorship, transparency, intellectual property and plagiarism exhibit lower publication metrics, hinting at potential gaps in comprehensive ethical investigations and warranting more scholarly focus. Furthermore, since LLM technologies are still evolving, there should be continuous ethical scrutiny and adaptation. Regarding LLM-related ethical challenges, future studies should continually reassess and adapt ethical frameworks to keep pace with technological progress. This will help foster responsible and accountable development, deployment, and use of LLMs while mitigating potential ethical problems and societal harms.

For researchers, our work suggests several directions. One is developing dynamic audit tools tailored to LLMs—monitoring systems that can flag emerging issues (e.g. new forms of hate speech or clever prompt attacks) in real time, given the models' evolving use. Another is exploring interpretable and steerable LLM architectures to reduce opacity: techniques like modular or causal diagrams of model internals that could let us pinpoint why a harm occurred. Interdisciplinary collaboration should be intensified, bringing ethicists, social scientists, and domain experts into the AI development loop early (as discussed in Section 6.4). On the governance side, experimental compute governance regimes (Section 6.3) and robust incident reporting requirements could greatly enhance oversight if trialed and adopted. We also encourage the creation of benchmark suites that evolve with models (Section 6.5), so that evaluation keeps pace with new capabilities and risks.
In concluding, we stress that the ethical challenges posed by LLMs are distinct in their scale and complexity, and they demand proactive, multi-faceted action. Unlike earlier AI systems, LLMs introduce phenomena like fluent misinformation and interactive agent ecosystems that strain traditional governance approaches. The coming years (2025 and beyond) will test whether industry, academia, and policymakers can jointly shape a responsible trajectory for LLM development. By highlighting pressing harms and sketching mitigation pathways, we hope this taxonomy provides a foundation for that effort. The goal is not to hinder innovation but to steer it: with guardrails informed by past incidents and ongoing research, LLMs can be developed and deployed in ways that uphold human values and social well-being. Achieving this will require continuous iteration on the ideas presented here, as both the technology and our understanding of it evolve. The conversation must remain open, integrative, and evidence-driven—much like the dynamic audit framework we advocate for the LLMs themselves. In sum, responsibly harnessed, LLMs can be a force for good; but realizing that potential while restraining the harms is the grand challenge that lies ahead for the AI community and society at large.

**Conflict of interest:** The authors declare that the research was conducted in the absence of any commercial or financial relationships that could be construed as a potential conflict of interest. **Acknowledgements:** in accordance with MLA (Modern Language Association) guidelines, we note the use AI-powered tools, such as OpenAI's applications, for assistance in editing and brainstorming.

**Institutional Review Board Statement:** Not applicable.

**Informed Consent Statement:** Not applicable.